# BlockJack: Towards Improved Prevention of IP Prefix Hijacking Attacks in Inter-Domain Routing Via Blockchain

I Wayan Budi Sentana, Muhammad Ikram and Mohamed Ali Kaafar
*Department of Computing, Macquarie University, 4 Research Park Drive, Macquarie Park, Sydney Australia*
*i-wayan-budi.sentana@hdr.mq.edu.au, {muhammad.ikram, dali.kaafar}@mq.edu.au*

Abstract: We propose "BlockJack", a system based on a distributed and tamper-proof consortium Blockchain that aims at blocking IP prefix hijacking in the Border Gateway Protocol (BGP). In essence, BlockJack provides synchronization among BlockChain and BGP network through interfaces ensuring operational independence and This approach preserving the legacy system and accommodates the impact of a race condition if the Blockchain process exceeds the BGP update interval. BlockJack is also resilient to dynamic routing path changes during the occurrence of the IP prefix hijacking in the routing tables. We implement BlockJack using Hyperledger FabricBlockchain and Quagga software package and we perform initial sets of experiments to evaluate its efficacy. We evaluate the performance and resilience of BlockJack in various attack scenarios including single path attack, multiple path attack, and attack from random sources in the random network topology. The Evaluation results show that BlockJack is able to handle multiple attacks caused by AS paths changes during a BGP prefix hijacking. In experiment settings with 50 random routers, BlockJack takes on average 0.08 seconds (with standard deviation of 0.04 seconds) to block BGP prefix hijacking attacks. The test result showing that BlockJack's conservative approach feasible to handle the IP Prefix hijacking in the Border Gateway Protocol.

## 1 INTRODUCTION

Border Gateway Protocol (BGP)–also known as Interdomain routing protocol–is a path-vector protocol that regulates the connectivity and information exchange among Autonomous Systems (AS).[1] Based on data presented in APNIC Research and Development (AS 65000), there are currently almost 70,000 unique ASN seen in their BGP routing table (BGP-Potaroo, 2020). Each AS maintains a number of IP Prefixes (in short Prefixes) and domains assigned by the Internet Assigned Number Authority through Regional Internet Registries (RIR).

To hijack IP prefix of a benign router, an adversary router (or AS) advertises a (*fake*) IP prefix that belong to another router (or AS). When the adversary AS conducts prefix advertisement, BGP sends the prefix to all neighbours on the Internet. As a result, the traffic that supposed to reach the original AS, is then redirected to the adversary AS which results in unavailability of crucial Internet services. Besides adversarial attempts, a number of cases are also caused by the unintentional mistake of network administrator during the routing setup (Hope, 2020).

Most recent researches involving machine learning have been conducted to detect prefix hijacking such as in (Qingye Ding et al., 2016), (Wu and Feng, 2009) and (Li et al., 2014). The first step in these techniques involving training of machine learning algorithms using the data acquired from dumped BGP control or data plane during the normal and hijacking condition. Then the machine learning algorithms monitor the current BGP update message and decide whether the condition is normal or not. Detection approach suffers to imbalance and resemblance data. Only a small number of hijacked BGP update message available currently compared to the total number of data produced by routers involved in the Internet. Moreover, to differentiate the traffic during normal and in hijacking condition is very challenging to conduct, even for research that involving deep learning (McGlynn et al., 2019).

---
[1] AS is an independent network that comprises the Internet and each AS assigned a 16-bit or 32-bit unique number known as Autonomous System Number (ASN).

To tackled the drawback of machine learningbased detection, several techniques propose blacklist based approaches (Alshamrani and Ghita, 2016), (Alshamrani and Ghita, 2016), (Testart et al., 2019). These techniques essetnially create profiles of ASes to blacklist malicious ASes (or router) to prevent them from routing advertisements. However, these approaches are generalized where certain router involved in malicious activities as a burden of the whole AS, whereas an AS can consist of hundreds of routers and can advertise hundreds of prefixes. Moreover, defining the threshold of secure and insecure ASes would be very challenging to conduct because almost all ASes may potentially involved in malicious activities. In order to secure BGP from prefix hijacking, the drawbacks exhibited by detection-based research makes prevention or mitigation approach are worth to explore.

In this paper, we propose BlockJack, a Blockchain-based model, to tackle the drawback of hierarchical RPKI model (Iamartino et al., 2015). In BlockJack, the pair IP prefix-ASN data and credential of each consortium member is stored in each Blockchain node to eliminate the need to resort to centralised or hierarchical, public key infrastructurebased schemes (Iamartino et al., 2015). BlockJack does not require to change the original BGP protocol; instead BlockJack provides a *Dispatcher* (see § 3.3 and Figure 1 for details) to automatically send filter commands to the router reducing software and hardware updates. Moreover, our proposed scheme is portable allowing non-BlockJack routers to communicate with BlockJack routers. The main contribution of this paper can be sumarized as follow:

- We present "BlockJack", a Consortium Blockchain-based model to verify prefixes and AS origin and also to neutralize the Prefix hijacking in BGP.

- We also propose a mechanism to increase the resiliency against AS Path changing and BGP routing Divergence during the Prefix Hijacking Neutralization.

- To foster further research, we release the source code of BlockJack and our experiments to the research community at: https://github.com/budisentana/prefixHijackingPrevention.git

## 2 BACKGROUND AND OVERVIEW

In this Section, we briefly present the several shortcoming of RPKI, reviews key basics of Blockchain, and highlight requirements of BlockJack system.

### 2.1 Resource Public Key Infrastructure

Internet Engineering Task Force (IETF) provides Resource Public Key Infrastructure (RPKI) to secure BGP against IP prefix hijacking. To prevent BGP hijacking, IETF releases RFC 6482 for Route Origin Authorization (ROA) and RFC 6483 for Route Origin Verification (ROV). ROA is a process where an AS authorizes a number of prefixes to be advertised under its jurisdiction, and stores it a tuple of IP prefix, AS (owner), the maximum length of AS and expiry date of each IP block (Iamartino et al., 2015). To prevent prefix hijacking, the tuple can be utilized during ROV process to verify whether or not an AS advertise the authorized prefix. Despite the crucial role, only 6.5% of total prefix announced in BGP are covered by ROA(Gilad et al., 2017) and there is no exact number of ROV coverage because of the invalid prefix in ROA is undefined(Hlavacek et al., 2018). 71% of the Internet Service Providers (ISPs) avoid to add more cost on the RPKI implementation and rely on its security using best practice mitigation by route filtering (Sermpezis et al., 2018). Moreover, RPKI hierarchical securing model potentially deteriorates of downstream Resource Certificate (RC) error during upstream RC overwrite or miss-configuration(Cooper et al., 2013).

RPKI architecture gradually delegates the RC from IANA as the global regulator to Regional Internet Registrar (RIR), ISP and private network company (Liu et al., 2016). Each of these institutions is allowed to publish a certificate of authority to its downstream and keep the RC in its storage. As suggested in (Cooper et al., 2013), any attack or missconfiguration on upstream network results to failure in prefix announcement for the downstream network.

### 2.2 Blockchain

Blockchain is a distributed data structure containing transactions of records forming a chain or blocks which are controlled by multiple Blockchain nodes. To provide data integrity, each block of transactions on a Blockchain has an individual digital signature created using a combination of the latest block's digital signature and a new digital signature, known as a digital footprint. Blockchain is also known as a distributed ledger that is completely open to any and everyone on the network. Blockchain allows all the network participants to reach an agreement (or consensus) during its operation.

The rapid development of Blockchain technology impacted the rise of various new versions of Blockchain models (Casino et al., 2019):

- Public Blockchain: A common feature of the public Blockchain is the need for miners to add new blocks to the existing block chain. Since the ownership is public, the identity of the accessing party can be anonymous (pseudo anonymous), it does not require permission (permissionless) to access the Blockchain system.

- Private Blockchain: This model is usually used by organizations with a centralized structure where each user is identified (permission) thus creating a trusted system or environment. The process of adding blocks to the existing Blockchain is carried out by the assigned *leader node*. Consensus mechanisms are widely used to prevent collisions from transactions running in parallel.

- Federated or Consortium Blockchain: This Blockchain model is semi-centralized, where the decision to approve each transaction is decided by the consortium members. Each consortium member needs permission to access the Blockchain. The process of adding blocks to the Blockchain is carried out by a leader who is chosen by the consortium members based on a consensus algorithm.

### 2.3 Requirements of BlockJack

We are using conservative approach in the BlockJack by keep the Blockchain and the BGP running in independent environment. This aproach preserve the BGP protocol so none of upgrade need by the router. For the data stored in the Blockchain, BlockJack only save the Prefix and its AS origin to create a resilient system against dynamic AS path changes and route divergence in the BGP network. In the following, we highlight two main requirements for BlockJack system:

Independent Environment: Instead of residing the node of Blockchain node in the router machine as in (Liu et al., 2019), we keep the Blockchain and routing environment running independently. As a bridge for those two environment, Profiler was created to serve the HTTP(S) request from Dispatcher that run a number of routine task to monitor BGP routing table. We have two consideration why we choose to create inter operable module between Blockchain and Router, that are:

- *Respecting legacy system:* Residing Blockchain node inside of router machine can change the role of dedicated router into multi purpose machine. Some update to the BGP protocol also needed to accommodate the Blockchain system. This approach will be so challenging to adopt in current condition relying on the result showing by(Sermpezis et al., 2018). Residing Blockchain inside the router machine can also raise a compatibility issue for the existing system.

- *Avoiding race condition between Blockchain access and BGP message interval:* BGP uses message signal–consists of tuple: OPEN, UPDATE, NOTIFICATION, and KEEP ALIVE–to periodically update the routing tables of routers. So if we reside the Blockchain node inside of the BGP protocol loop, we need to use one of those message to trigger our Blockchain system. This condition force all the process that accessing the Blockchain, including prefix authorization or verification, should be completed before the next interval of the BGP message. The race condition occur when the Blockchain process exceed the BGP message interval. Related to Prefix advertisement, BGP uses UPDATE message to find a new prefix advertised, withdrew or updated from its Peer or neighbors. In default, the Minimum Advertisement Route Interval (MARI)(Liu et al., 2020) is set every 30 seconds. It implies that all the prefix authorization, which is adding a prefix to the Blockchain, should be completed within 30 seconds. This could be challenging due to complex process of consensus mechanism in Blockchain.

Resiliency Against AS Path Changing: For routing operations, BGP protocol uses three tables: *i) BGP Neighbor Table*: containing information about BGP neighbors, *ii) BGP Table* (BGP RIB or BGP topology table): contains the list of prefix (network) and its routes (several routes are directed to the same network with different attributes), and *iii) BGP Routing Table*: contains selected *valid-best* routes from BGP Table. BGP table learns the route to the source of the prefix from its neighbors. Each route in

BGP table list consists of several attributes including local pref, AS path, MED (Multi Exit Discriminator), and next hop (Attarde and Dhamal, 2009).

BGP uses the values in these attributes to decide the *best* and *valid* path to get to the source of the IP prefix. The chosen route is then stored to the BGP routing table which is then announced to the immediate neighbors. The *best-valid* path can dynamically change as the value of those parameters change. If there is a change in the *best-valid* status of a route in a router, the router will send an BGP message UPDATE to its peers. This change is potentially affecting the routing table on neighboring routers. If the AS path changing is oscillating then it is known as BGP routing divergence. Routing divergence can be caused by load balancing policies and also BGP routing policies(Ahmed and Sarac, 2014).

Saad et al., (Saad et al., 2019), leverage AS paths, stored in blockchain module, as the main parameter to detect prefix hijacking. Most of the data is used only by the router itself to verify if the same AS has more than one routes. The study assumes that the path to reach the prefix source is fixed and captured in convergence routing conditions. However, this approach is not valid if there is traffic redirection caused by a malfunctioned AS as a result of network failure. As its nature, the router will look for alternative paths in case of network failure so that data exchange can continue. If there is a change in the path, the valid-best status on the previous path is likely to change. As a result, the AS path in the latest condition with valid and best status will be different from the AS path stored on the Blockchain so that all prefixes announced by the AS experiencing valid traffic redirections.

To accommodate AS route path-changing and divergence, BlockJack stores only prefix and its AS source (origin) in the Blockchain. BlockJack uses AS path only to retrieve the prefix source and discharge the rest of it. As a complement, BlockJack uses next hop information to identify the peers that contribute to the addition of prefix in the routing table, in order to create the *Inbound filter* when the hijacking occur. Inbound filter is a function use to create a filter for the incoming Prefixes or AS from a certain AS through the immediate neighbors. This approach reduce the number of verification and authorization process to the Blockchain caused by dynamic change of AS path.

BlockJack only triggered if there is any changes in valid and best path status and change of the next hop. To accommodate this approach, BlockJack prepares two main features, namely prefix authorization, which is used by AS prefix owners to claim or authorize prefixes, and prefix verification which is used by the whole AS to verify prefixes received from its neighbors. We provide more details about these two feature in § 4.1 and § 4.2

## 3 SYSTEM ARCHITECTURE

Figure 1 depicts the three modules of BlockJack: Blockchain, Profiler and Dispatcher. The Blockchain module handling the Smart Contract, Certificate of Authorization (CA) provider, data storage (Ledger), and a consensus mechanism while the Profiler creates routers' profiles and facilitates a gateway to the Blockchain ledger. It also also provides wallets to store all router credentials under a certain AS authority. The Dispatcher module conducts routine tasks to monitor the routing tables and dispatches filtering commands if there is any update on BGP routing tables.

### 3.1 Hyperledger Fabric Blockchain

We leverage Hyperledger Fabric [2] platform to build the Blockchain module. Unlike public-based (permissionless) Blockchain, Hyperledger fabric eliminates the role of miners in tethering the new blocks to the existing blocks. The consensus mechanism assures only trusted and known consortium member parties can be involved in Blockchain transactions. Hence, this Blockchain model is appropriate to regulate the interaction among Autonomous Systems that demand a highly trusted environment.

Replacing the role of miners and other public based consensus, Hyperledger Fabric handled its consensus mechanism by adding several components including Orderer, Endorser, Chaincode and the Ledger itself(Linux-Foundation, 2020b). For this research we modify the Chaincode and align the Ledger structure as needed by the BlockJack. Chaincode is a code of a program that handles the business logic of the transaction among consortium members. Chaincode acts as a smart contract and used by Endorser as a matrix to approve or disapprove a transaction. While, Ledger is a database that stores all the transactions approved by consensus mechanism and provides the query access for an authorized entity.

---

[2] The Hyperledger Fabric is a consortium based Blockchain that only allows identified entities (permissioned) to access the network(Linux-Foundation, 2020b).

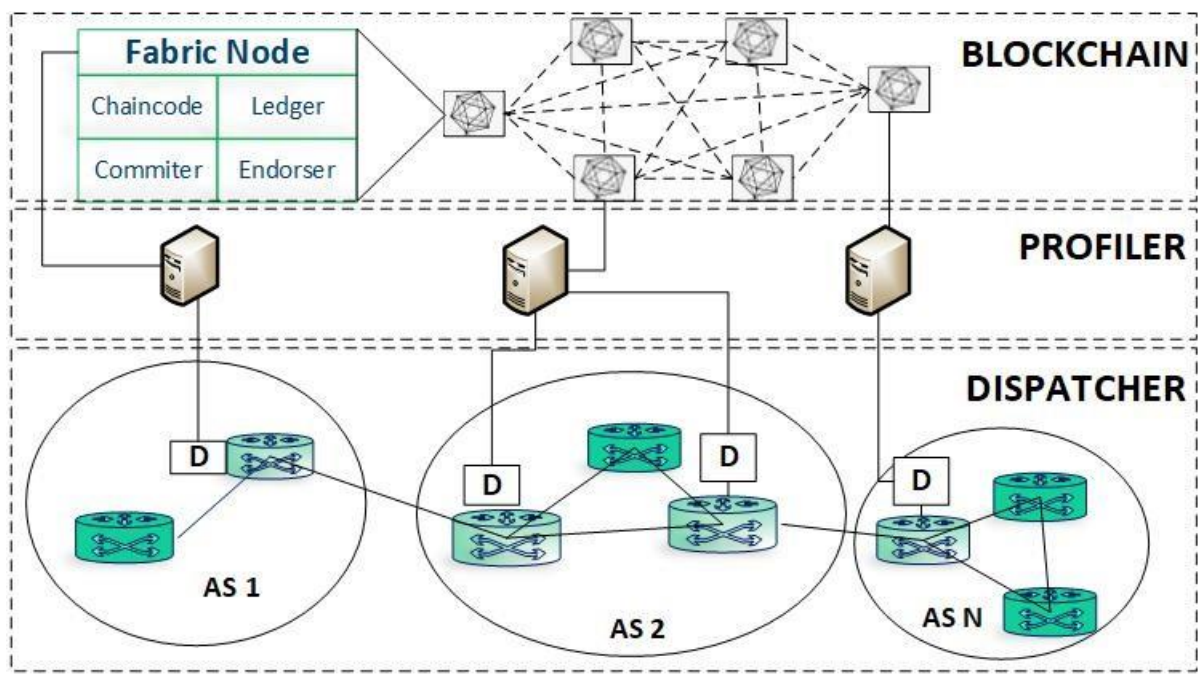

Figure 1: An overview of the three modules of BlockJack. Blockchain module is handling data storage and data query while Profiler is used as a bridge between Blockchain and Dispatcher as well as to store the Credential of each router. Dispatcher *monitors* routers and *dispatches* filter commands if Prefix Hijacking occurs.

In the Hyperledger Fabric, ledger consists of two different but related parts, that is World State and Blockchain(Linux-Foundation, 2020b). World State is a database that stores current values of a set of ledger states. It is allowed a transaction request to access the current value directly without need to traverse the value in the whole transaction log. The world state may change dynamically when a transaction states created, updated or deleted. While, Blockchain records all the changes appears in the current world state and stored in transaction log. Every time the commit order occurs, the transactions in the world state are collected inside the block and then appended to the Blockchain. Hence, the Blockchain consists of changing history resulting in the current world state that cannot be modified.

For this research purpose, we prepare the ledger to correspond to the Chaincode plus transaction key that is taken from the corresponding prefix so the retrieval process becomes faster due to indexing process. We are creating four columns of table that consist of PREFIX, ASN, DOCUMENT TYPE and ACTIVE STATUS of the prefix. Active status is needed if the temporary withdrawal occurs in the BGP table, hence the re-announcement of the prefix does not create a new transaction and only change the status of the prefix

### 3.2 Profiler

Profiler is used as an interface between the Dispatcher and the Blockchain. Figure 2 shows Profiler module consist of three parts that are Admin function, router profiler function and Rest API function. Admin function used to create a credential for the administrator before the admin can create a router profile, by invoking the Fabric CA module in the Blockchain.

Router Profiler is a function that can be used by administrators to create profiles of each router from its internal AS. The profile consists of router-id and the AS number. The router-id is then used as a username for the router to be sent to fabric CA to create the router certificate of credential. All the credentials of admin and routers are stored in the wallet. And the last function of this module is a Rest

API server that can be used as a gateway from router dispatcher to the Blockchain.

Rest API Server provides functions to add prefix into the Blockchain and query prefix from the Blockchain. Respectively, those functions are useful for the prefix authorization and verification pro-cess. This function also provides an authentication routine for each http request from the router and equips the request with the corresponding router credential before it is sent to the Blockchain.

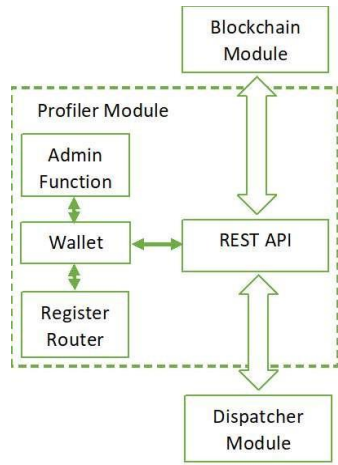

Figure 2: Overview of *rest* API and router *profiler* module. Admin and router profiler function is used to create credentials of administrator and router. The credentials are then stored in the wallet and used by the rest API server to equip the request from Dispatcher to access the Blockchain module.

### 3.3 Router Dispatcher

Router Dispatcher (Dispatcher) is used to interact with router machines. Although it works in the router, the routine inside Dispatcher is independent of BGP routing signals. This approach allowed regular routers to connect to routers that equip with BlockJack, hence minimizing the update for the router software. Dispatcher consists of three routines which are monitor, sender and verifier, as shown in Figure 3. There are also two local caches that are *Local ROA*, used to store temporary prefixes announced by router in internal BGP mode and *Local ROV* that use to store temporary prefixes announced by neighboring ASes.

Monitor routine observes the BGP routing table while Monitor routine consists of *shell script* which sends Secure Socket Layer (SSL) command to query BGP routing table. The command returns the value of *network (pefixes), next hop, metric, local preference, weight, and AS-path* for each route available to get to the source of the prefixes. From these return values, BlockJack takes the *network, next hope and AS-path*. Monitor then split the *AS-path* to find the origin AS of each prefix, discharge the rest of AS in the *AS-path*, and then combine it with *prefix* and *next hope*. If the prefixes originate from internal AS then the combination value is assigned to *ROA variable*, otherwise it is assigned to *ROV variable*. As explained in § 2.3, Monitor only captures the route with status *valid and best* to reduce the authorization and verification process to the Blockchain.

Sender routine assists the Prefix Authorization process in § 4.1. Sender consists of functions to compare *ROA variable* generated by Monitor routine with *local ROA cache* to seek any announcement or withdrawal of prefixes from internal AS origin. If a new prefix announcement is found, Sender then sends HTTP(S) requests to the Profiler to add a new prefix and its origin to the Blockchain. When a prefix withdrawal occurs, Sender sends a request to update the prefix status in the Blockchain.

The Verifier routine supports the Prefix Verification as in detailed in § 4.2. The main function of the Verifier routine is to send the new prefixes announced by the router's neighbors for verification against the data stored in the Blockchain. Verifier uses the *ROV variable* provided by Monitor and compares it with the *Local ROV cache* to find new prefixes announced by the neighbors. Verifier also consists of a function to send Inbound filter commands to the router to restrict an incoming prefix from certain neighbors when the prefix is identified to become a source of collision in the routing table.

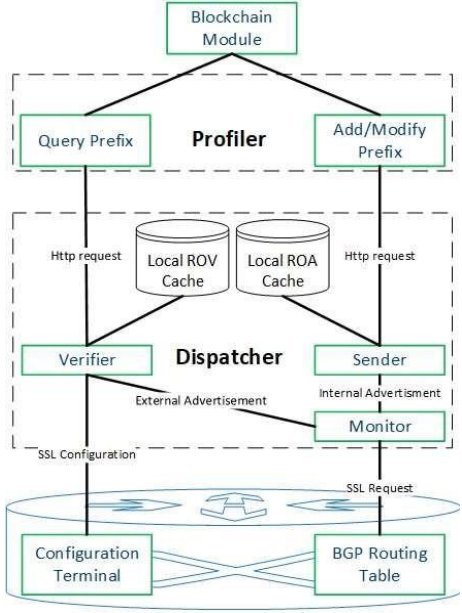

Figure 3: Overview of *Dispatcher* module where Monitor routine observes the BGP routing table and finds the prefixes announced by internal and external ASes. Internal prefix list will be stored in the Blockchain using Sender routine while external prefixes will be verified to the Blockchain using Verifier.

## 4 PROCESS AND MECHANISM

BlockJack consists of a supporting mechanism which includes Blockchain initialization, register admin and router, and main mechanism which includes prefix authorization and prefix verification. Supporting mechanism is the process used to operate the Blockchain network and to create credentials for admins and routers who will interact with the Blockchain. In the following, we provide more details about the *main* mechanisms in BlockJack:

### 4.1 Prefix Authorization

By leveraging Blockchain, ASes uses Prefix Authorization to prove the ownership of their prefixes. Prefix Authorization starts when the Dispatcher module reads the BGP routing table. Monitor routine in the Dispatcher, reads and defines the prefixes route announcements by the internal AS router, and saves it in a variable. Monitor captures the routes that appear with the status of *valid and best* to reduce the number of prefixes list. The Sender routine then compares the values resulting from Monitor to the local ROA cache to find a new prefixes announcement. If a new prefix is found, the Sender routine will send a https request to the Profiler asking for prefix addition to the Blockchain. In the Profiler, the Rest API server receives the request and authenticates the request to the router profile. Once it authenticates, the Profiler adds the router credential to the request and sends it to the Blockchain.

In the Blockchain, Endorser verifies the request based on the contract written in the Chaincode. The smart contract checks whether the new prefix is violating the same prefix previously stored in the Blockchain. Once the request complies with the smart contract, the Endorser sends the prefix to the Orderer and then the consensus mechanism in the Blockchain is started. The Orderer distributes the request to all consortium members to have approval from at least 50% of the consortium members. When the consensus is achieved, the Orderer signaling the committer to seal the transaction and the Orderer will add a new block into the ledger (Blockchain). At the end of the Blockchain transaction, the Orderer creates a new *World State* (see § 3.1) and distributes it to the consortium members to update the ledger state.

As prefix addition and modification processes invoke a consensus mechanism in Hyperledger to add a new block into the chain, they both are costly to be conducted by the BlockJack model. As shown in (1), the cost to authorize prefixes $x$ in $C_A(x)$ is equal to the latency (*delay*) of requesting it through Profiler in $L_R(x)$ plus the total latency to have Endorsement approval from at least 50% of each consortium member (*i*) in $L_E(x)$ and also the latency to create a new block and distribute it among $N$ consortium member in $L_D(x)$. We omit the delay used by Committer to seal the transaction since it is only a simple process in the Blockchain. Although the Prefix Authorization process is costly, the total cost of adding prefixes into Blockchain can be spread among the prefixes owner, since each AS only needs to authorize their own prefixes.

$$C_A(x) = L_R(x) + \sum_{i=1}^{\frac{N}{2}}(L_E(x)) + \sum_{i=1}^{N}(L_D(x)) \quad (1)$$

### 4.2 Prefix Verification

Prefix Verification aims to check whether the announcement received by the router from its neighbors contains the authorized prefix. For this purpose, BlockJack compares the prefixes found in the routing table with the prefixes stored in the Blockchain. Prefix Verification mechanism started in the Dispatcher Module. The monitor routine captures the prefixes route with status *valid-best* and assigns the prefixes announced by the neighbors into *ROV variable*. The verifier routine then compares this variable with the prefixes stored in the Local ROV cache to find a new prefixes announcement. If a new prefix is announced, the Verifier requests for verification to the Blockchain through Profiler. In the Profiler Module, the request is accepted by the Rest API server, and after authenticating against the router credential, the request is then sent to the Blockchain.

In the Blockchain module, the Chaincode verifies the compliance of the request to the Smart Contract and then query the Ledger. The query result then verifies by Smart Contract and publishes three kinds of signals, which are valid, invalid, or unknown. The valid signal indicates that the prefix sending to the Blockchain is available and corresponds to the right AS Number. While, an invalid signal denotes that the prefix exists in the Ledger but corresponds to the different AS Number. This invalid signal can indicate the BGP prefix hijacking occurs in the network. The last unknown signal indicates neither the prefix and the AS Number available in the Ledger. For this research, the unknown signal is to return a valid signal for Dispatcher to allow non-BlockJack routers to announce their prefixes and to reduce the false-positive number of verification. When the Verifier accepts the invalid signal, then it produces Inbound filter commands to block the announcement from the AS that indicated as a source of prefix hijacking. The Inbound filter needs a parameter of *next hop* that can be found in the ROV variable as it is assigned by the Monitor routine before the verification.

Compared to the authorization mechanism, prefix verification is straightforward: after the router complies with the credential and smart contract, data retrieval can be conducted in the ledger without any approval of other consortium members. Moreover, the ledger retrieved for verification is resided in the correspondence Blockchain node and more specifically in the current *World State* ledger (see § 3.1). Hence the verification process can be much faster than authorization.

# 5 TESTBED SETUP AND EVALUATION

This section presents, our experiment testbed and analysis.

## 5.1 Experiment Testbed

Blockchain module of BlockJack is implemented using *Hyperledger Fabric* (Linux-Foundation, 2020a). Each Blockchain node, including *Orderer, Fabric CA,* and *Chaincode* are running in a separated Docker container (Docker-Inc, 2020). For the Profiler module, we tailor *Node js* to create the admin function, router registration, and Rest API server. We built the Dispatcher module based on *Python* and combined it with *shell script*. BGP network environment developed using *Quagga* router software, and for this research, we are using *Dockerized Quagga image* produced by (Chiodi, 2020). Each router running in a separate *detach-mode* Docker container connected by a customized virtual network. We use the testbed and conduct several sets of experiments to evaluate the performance and resiliency of BlockJack against BGP hijacking attacks. We provide further details in the following:

Performance Evaluation: We aim to evaluate the performance of our BlockJack in terms of the time processing for handling prefix authorization and prefix verification requests. To this end, we conduct two sets of experiments [3] and record the processing time of the authorization and verification mechanism of BlockJack. In particular, we generate sets of random prefixes from a BlockJack router and query the Blockchain ledger to determine the authorization and verification times required by our proposed systems.

For the prefix authorization setup, we create a function in the Dispatcher module to send a various number of prefixes into the Blockchain to measure BlockJack performance in handling prefix authorization. Each prefix authorization process followed by *commit* order so each prefix addition will be adding a block in the Blockchain. At the end of the experiment there will be a thousand blocks in the chain. To measure the prefix verification time, we create a script so the Dispatcher can send a various number of prefixes to be verified by Blockchain with 1000 blocks. Resiliency Evaluation: Resiliency evaluation aims to observe BlockJack resilience in neutralizing prefix hijacking attacks with various scenarios. For that purpose, we leverage Quagga and Docker in the higher computing environment[4] and create several network topologies with various number of routers of 20 to 60. We measure the prefix hijacking neutralization in two stages including Prefix Prepending and Neutralization. Prefix Prepending is the process of adding an ASN to the AS-path parameter in the BGP table for each AS passed by a prefix. In this experiment, the prepending time is equal to the time needed by the adversarial prefix to arrive at the router where the Dispatcher resides (BlockJack router) and disrupts the original prefix as the route with the *valid-best* status. While, Neutralization is the stage where BlockJack detects, verifies, and sends filter commands to the router to neutralize hijacking. By measuring the prepending (BGP hijacking attack) time and neutralization (blocking) time, we determine the duration of BGP hijacking attacks and the efficiency of BlockJack to neutralize the attacks, respectively.

---

[3] Experiments were performed on a workstation with Ubuntu 18.04, CPU 2.7GHZ, and RAM 16GB.

For this experiment, we create three different scenarios as describes bellow.

- Single Path Attack Scenario. This first scenario aims to evaluate BlockJack resilience in neutralizing attacks originating from a single path. We create a binary tree-like network topology and reside the dispatcher on the router which is located at the root. We prepared five adversarial prefixes that would be used by routers located in the farthest branch to hijack the prefixes announced by routers in the leaf of the tree. These attacks create single paths when they reach the BlockJack router (root).

  We prepared five experimental sets in various router numbers of 20 to 60, with five experiments in each set and recorded the Prepending and Neutralization times for each trial. To give a fair treatment to each experiment set, we restart the Blockchain network for each experiment setup. This will make the Blockchain ledger only consist of the genesis block in each time the experiment is started.

- Multiple Path Attack Scenario. This scenario is designed to examine BlockJack's resilience in anticipating routing path changes that occur during BGP prefix hijacking as explained in 2.3. In this scenario we modify the binary tree network topology in the first scenario by setting up BGP peering for each branch at the same level. This will cause each announced prefix to have more than one path when it comes to the BlockJack router that is at the root of the tree.

  Similar to the first scenario, we also created a script to bypass the prefix authorization and prepared a total of 25 experiments on various setups with the number of routers varying from 20 to 60. We also restarted the blockchain network to get fresh genesis blocks for each experiment set.

- Random Attack Scenario. This scenario is set up to examine the BlockJack resilience in a very random BGP environment. We made several random network topologies with various numbers of routers (20 to 60). The connectivity level in each experiment was set to 25% indicating that a node has a probability to be connected to 25% of the total nodes on the network, except for the experiment with the number of routers of 20 which we set at 50% to avoid the occurrence of unconnected node in the network. We also created a script to place the Dispatcher on a random node (router) and prepared 5 random adversarial prefixes which are used to hijack the prefix announced by the original router in each set of experiments.

  We ran BlockJack for 10 minutes for each experiment and tried to get 5 sets of experimental results on a different number of routers. Since the dispatcher and adversarial prefixes are randomly assigned, it is possible that the hijacking process will not affect the router where the dispatcher is placed. In that case, we will discard the experiment results and run it again until we find that at least one prefix in the routing table is affected by hijacking. We also assume that if within the specified time period the adversarial prefix does not appear in the routing table, then we assume that the hijacking

---

[4] Experiment conduct in Cluster server with 4 core CPU units, 128 GB of memory, 500 GB hard drive and leveraging Ubuntu 18.04 LTS.

process does not affect the routing table where the dispatcher is located.

*Limitation:* Although we managed to run 70 routers on the docker container, we failed to configure the router running on top of the container using the telnet command. Problems with overloaded networks give rise to a network request timeout error message. Docker network shares the same Linux kernel to handle all virtual networks running on top of the Docker container. This will cause an overload on the virtual network.This condition also causes the Hyperledger failed to install the Chaincode (smart contract) on the Blockchain node through the corresponding port.

## 5.2 Analysis

Performance Analysis: The measurement results of the average and total time required to perform Prefix Authorization and Prefix Verification are presented in Table 1. The number of prefixes in the table is proportional to the number of blocks in the Blockchain, considering that each prefix addition is always followed by a commit command which means adding a new block to the blockchain. More details about the results of the prefix authorization analysis and prefix verification are presented as follows:

Table 1: Time recorded by BlockJack to *authorize* and *verify* prefixes. Each prefix addition followed by commit process and create a new block.

| # of Pref. (Block) | Authorization | | Verification | |
|---|---|---|---|---|
| | Avg. (s) | Total (s) | Avg. (s) | Total (s) |
| 100 | 2.16 | 216.21 | 0.1 | 10.27 |
| 500 | 2.15 | 1,076.61 | 0.09 | 47.38 |
| 1,000 | 2.15 | 2,154.32 | 0.09 | 92.12 |

Prefix Authorization Analysis: Table 1 shows prefix authorization time increase gradually according to the number of prefixes sent to the Blockchain. BlockJack needs 2,154.32 seconds to authorize 1,000 prefixes with an average authorization time of 2.16 seconds, which is quite heavy even for a small two node Blockchain environment. This expensive process is caused by a complex consensus mechanism during the block addition to the chain as described in § 4.1.

Regarding BGP messages set up by Cisco router (Vinit and Brad, 2018), if the BGP UPDATE interval is 30 seconds, the maximum prefixes that can be authorized in that interval is 13 prefixes. These 13 prefixes per interval are higher than 12 prefixes average announcement per AS origin as shown in (Tony Bates, 2020). Hence, in this case, BlockJack can handle the average prefixes announcement in real world condition in one BGP UPDATE message interval simultaneously without the assistant of the Local ROA cache, disregarding the network traffic delay.

Since we adopted a conservative approach for the BlockJack, we calculate the prefix authorization time for the case of the highest number of prefix ownership by an AS. According to (Tony Bates, 2020), the highest number of prefixes announced by an AS is recorded by AS8151 (Uninet S.A. de C.V., MX) with 8125 prefixes. In this case, BlockJack needs 625 BGP UPDATE interval or 17,550 seconds to authorize the whole prefixes when it runs for the first time. That result is clearly created race condition between BGP UPDATE interval and prefix authorization if the Blockchain node resides in the router machine and depends on BGP message signal. The time taken to access Blockchain during the authorization process far exceeds the BGP UPDATE message interval. Hence, Blocjack approach by keeping the Blockchain and routing environment running independently is reasonable considering the real-world condition.

Prefix Verification Analysis: The experiment result in Table 1 shows that the prefix verification process is much lighter compared to prefix authorization. BlockJack needs 92.12 seconds to verify 1000 prefixes or on average of 0.09 second per prefix. Given the worst scenario that the growth projection of 150 prefixes per day(Geoff, 2020) appear in the concurrent time, and assume that the same amount of prefix withdrew in the same time, then BlockJack only need 27 seconds to verify the 300 updated prefixes. This is below the 30 seconds BGP UPDATE message interval. As an addition, none of the consortium members growth will affect the verification time because the process only need to query the internal ledger without consortium members' approval. Moreover, Hyperledger Fabric provides a *World State* so that the BlockJack does not need to trace all log transactions to perform data retrieval. Hence, in this case BlockJack can handle Prefix verification without Local ROV cache assistance.

However, verifying the global BGP routing table would be so challenging to conduct in one BGP update message interval. Expect that the number of the global routing table is 850,000(BGP-Potaroo,

2020) and assume that the verification time per prefix is 0.09 seconds, then the total time needed by BlockJack to verify those prefixes on the first time running is 76,500 seconds. That result is equal to 2,550 of the BGP UPDATE message interval. Hence, the existence of a local ROV cache is crucial to reduce the verification request during the BlockJack operation. The Dispatcher can compare the entry of BGP routing table and the local ROV cache to find a new prefix announcement or withdrawal occurrences, and then verify those prefixes. Moreover, any verification that exceeds the BGP message interval is not impacting the BlockJack because the Dispatcher is not dependent on any BGP message signal as described in § 2.3.

Resiliency Analysis: BlockJack is able to neutralize all adversarial prefixes that disrupt the BGP routing table. BlockJack ignores the adversarial prefix under the following conditions; first, an adversarial prefix which does not take over the original prefix to be the valid-best path. This is possible when the position of the original prefix is closer to the adversarial prefix or the accumulation of all attribute values of the original prefix in the BGP table is better than the adversarial prefix; second, when the position of the AS path adversarial prefix overlaps with the other adversarial prefixes. This will cause the adversarial prefix with the longer AS path to be automatically neutralized when the adversarial with the shorter AS path is neutralized.

The results of BlockJack's resilience evaluation are depicted in Figure 4. We observe that the average prepending time increases gradually in accordance with the router addition in single path and multiple path attack scenarios, while the average prepending time for random attacks seems to fluctuate. The lowest prepending times for each single path, multiple path and random attack scenario were 28.068, 41.855, and 52.101 seconds, which are recorded during the experiment with 20 routers. The average prepending time for all experimental sets on single path, multiple paths and random attack scenarios, respectively, is 74.527, 80.9088, and 54.9572 seconds.

Neutralization time looks constant in all scenarios with an average of 0.1516 seconds in single path scenarios and 0.2362 seconds in multiple path scenarios, except for random attack scenarios which record an average neutralization time of 1.0484 seconds. The amount of neutralization time in random attack scenarios is up to 5 times compared to multiple path scenarios. This is because the number of neighbors in the random attack scenario is greater than multiple path scenarios. From the five adversarial prefixes sent, the average number of attacks on the random scenario reaches 10.08 attacks, compared to 4.52 attacks received by BlockJack routers in multiple path attack scenarios. If we take a sample of random attack scenario with 50 routers, on average BlockJack needs 0.957 seconds to neutralize 12.04 attacks of five experiment attempts. That record is equal to 0.08 seconds to neutralize a single attack.

The standard deviation of prepending and neutralization time is also seen constant in single path and multiple path scenarios with an average range of 8.2488 seconds to 11.6154 seconds for prepending time and 0.0162 seconds to 0.0188 seconds for neutralization time. While the average standard deviation of prepending and neutralization time in the random attack scenario was recorded at 20.3712 seconds and 0.8998 seconds, respectively.

# 6 Related Work

Previous research involving Blockchain as a source for prefix validation has been proposed by (Xing et al., 2018). This paper introduces BGP Coin to be used as Capital in prefix validation and built on the Ethereum platform. The idea is mimicking cryptocurrency model whereas every entity wishes to check the prefix validity should provide a number of Ether (coin) to mined so the miner can attach a new prefix block to the chain. The similar concept with BGP Coin is proposed by(Paillisse et al., 2018) that proposes IP Chain. This research assumes the prefix shares the same characteristic as a coin in the Blockchain that can be allocated, transferred, divided and cannot be assigned to two participants at the same time. The system is built on Ethereum using Proof of Stake (PoS) consensus. Although PoS does not need miners, it needs Validator to attach a new block to the chain, where this validator needs to deposit an amount of Ether coin. Paper (de la Rocha Gomez-´ Arevalillo and Papadimitratos, 2017) proposed a general concept of Blockchain for prefix validation independent to consensus mechanism. However, this proposal lacked technical detail and emphasis on general requirements that should have been met by the system. As a part of protocol mentioning about the mining process, we assumed that this proposal was pre-pared for a public based Blockchain. All of the mentioned research was using a public based Blockchain that needs miners and coins during the transaction and becomes the deficiency of those systems in a closed environment such as prefix authorization and validation.

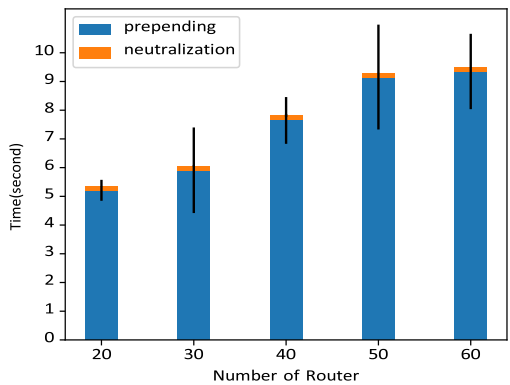

(a) Single Path Attack.

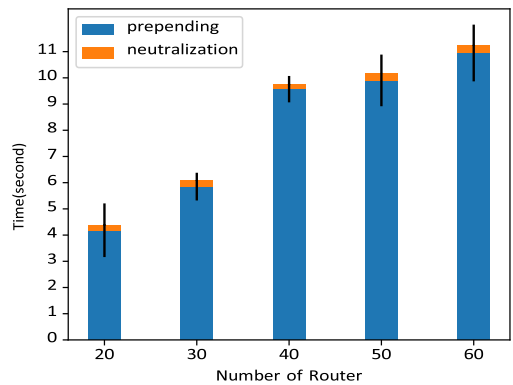

(b) Multiple Path Attack.

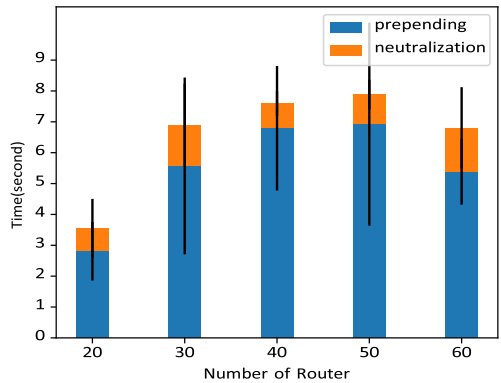

(c) Random attack.

Figure 4: Resiliency Evaluation result. The prepending time record increases gradually as the number of routers increases in single path and multiple path attack scenarios. While, the prepending time for random attacks fluctuates. All prepending times are presented in the form of time / 10 seconds.

The research involving private-based Blockchain to secure BGP routing was introduced by (Saad et al., 2019). This research tailoring Proof of Authority (PoA) in Ethereum platform. Unfortunately, this paper only emphasizes the Blockchain side and neglects the network protocol. The simulation of the BGP securing process only conducted by simulation. The next research involving private based Blockchain is shown in (Liu et al., 2019) and (Liu et al., 2020). These are two related papers that tailoring Hyperledger Fabric platform and Quagga router software. However, this research conducts a modification to the BGP protocol in Quagga router software. Moreover, this research deploys the Blockchain node inside of the router software. This approach would be so challenging to adopt in the real-world case because the router should be installed with all Blockchain environment prerequisites including Docker, Node js, and other programming languages needed to create a smart contract. That would change the router role to become a multi-purpose machine.

The closest concept to the current condition of BGP routing was conducted by (Sfirakis and Kotronis, 2019). This paper introduced the concept of a Blockchain-based prefix hijacking prevention in technical network detail. The concept was deployed in the Bitcoin platform and Quagga router software. Unfortunately, this research is still tailoring Bitcoin that needs a miner for the block execution.

## 7 CONCLUSION

Although the Prefix Authorization and Prefix Verification processes can ideally be handled in one BGP Update message, several conditions will cause a race condition between processes that occur on the Blockchain and processes that occur in BGP. Therefore the conservative approach adopted by BlockJack is right to deal with the problem. BlockJack also manages to handle dynamic-multiple hijacking due to changes in the BGP attribute value which causes dynamic changes in determining the best-valid path in the BGP routing table.

## REFERENCES

Ahmed, N. and Sarac, K. (2014). Measuring path divergence in the internet. In *2014 IEEE 33rd International Performance Computing and Communications Conference (IPCCC)*, pages 1–8.

Alshamrani, H. and Ghita, B. (2016). Improving ip prefix hijacking detection by tracing hijack fingerprints and verifying them through rir databases. In *Proceedings of the 13th International Joint Conference on E-Business and Telecommunications*, ICETE 2016, page 57–63, Setubal, PRT. SCITEPRESS - Science and Technology Publications, Lda.

Alshamrani, H. and Ghita, B. (2016). Ip prefix hijack detection using bgp attack signatures and connectivity tracking. In *2016 International Conference on Software Networking (ICSN)*, pages 1–7.

Attarde, S. A. and Dhamal, S. K. (2009). Persistent bgp deviation. In *Proceedings of the International Conference on Advances in Computing, Communication and Control*, ICAC3 '09, page 86–91, New York, NY, USA. Association for Computing Machinery.

BGP-Potaroo (2020). Bgp analysis report-bgp table. https:// bgp.potaroo.net/index-bgp.html. BGP Table Data last accessed 1 July 2020.

Casino, F., Dasaklis, T. K., and Patsakis, C. (2019). A systematic literature review of blockchain-based applications: Current status, classification and open issues. *Telematics and Informatics*, 36:55 – 81.

Chiodi, P. C. (2020). Quagga router software code @github. https://github.com/pierky/dockerfiles/tree/master/quagga. last accessed 1 July 2020.

Cooper, D., Heilman, E., Brogle, K., Reyzin, L., and Goldberg, S. (2013). On the risk of misbehaving rpki authorities. In *Proceedings of the Twelfth ACM Workshop on Hot Topics in Networks*, HotNets-XII, New York, NY, USA. Association for Computing Machinery.

de la Rocha Gomez-Arevalillo, A. and Papadimitratos, P.˘ (2017). Blockchain-based Public Key Infrastructure for Inter-Domain Secure Routing. In Camenisch, J. and Kesdogan, D., editors,˘ *International Workshop on Open Problems in Network Security (iNetSec)*, volume IFIP eCollection-1 of *Open Problems in Network Security*, pages 20–38, Rome, Italy.

Docker-Inc (2020). Docker container install code @github. https://github.com/docker/docker-install. Last accessed: 18/12/2020.

Geoff, H. (2020). Bgp in 2019 – the bgp table. https://blog.apnic.net/2020/01/14/bgp-in-2019-the-bgp-table/. Blog APNIC last accessed 1 Juli 2020.

Gilad, Y., Cohen, A., Herzberg, A., Schapira, M., and Shulman, H. (2017). Are we there yet? on rpki's deployment and security.

Hlavacek, T., Herzberg, A., Shulman, H., and Waidner, M. (2018). Practical experience: Methodologies for measuring route origin validation. In *2018 48th Annual IEEE/IFIP International Conference on Dependable Systems and Networks (DSN)*, pages 634–641.

Hope, A. (2020). Russian rostelecom compromises internet traffic through bgp hijacking. https://www.cpomagazine.com/cyber-security/russian-rostelecom-compromises-internettraffic-through-bgp-hijacking. Accessed: 18/12/2020.

Iamartino, D., Pelsser, C., and Bush, R. (2015). Measuring bgp route origin registration and validation. In Mirkovic, J. and Liu, Y., editors, *Passive and Active Measurement*, pages 28–40, Cham. Springer International Publishing.

Li, Y., Xing, H., Hua, Q., Wang, X., Batta, P., Haeri, S., and Trajkovic, L. (2014). Classification of bgp anomalies´ using decision trees and fuzzy rough sets. In *2014 IEEE International Conference on Systems, Man, and Cybernetics (SMC)*, pages 1312–1317.

Linux-Foundation (2020a). Hyperledger fabric project code @github. https://github.com/hyperledger/fabric. Hyperledger Fabric Project last accessed 1 July 2020.

Linux-Foundation (2020b). Hyperledger fabric release 2.0. https://hyperledger-fabric.readthedocs.io/en/release-2.0/whatis.html. Hyperledger Fabric Release 2.0 readthedocs last accessed 1 July 2020.

Liu, X., Yan, Z., Geng, G., Lee, X., Tseng, S.-S., and Ku, C.-H. (2016). Rpki deployment: Risks and alternative solutions. In Zin, T. T., Lin, J. C.-W., Pan, J.-S., Tin, P., and Yokota, M., editors, *Genetic and Evolutionary Computing*, pages 299–310, Cham. Springer International Publishing.

Liu, Y., Zhang, S., Zhu, H., Wan, P.-J., Gao, L., and Zhang, Y. (2019). An enhanced verifiable inter-domain routing protocol based on blockchain. In Chen, S., Choo, K.-K. R., Fu, X., Lou, W., and Mohaisen, A., editors, *Security and Privacy in Communication Networks*, pages 63–82, Cham. Springer International Publishing.

Liu, Y., Zhang, S., Zhu, H., Wan, P.-J., Gao, L., Zhang, Y., and Tian, Z. (2020). A novel routing verification approach based on blockchain for inter-domain routing in smart metropolitan area networks. *Journal of Parallel and Distributed Computing*, 142:77 – 89.

McGlynn, K., Acharya, H. B., and Kwon, M. (2019). Detecting bgp route anomalies with deep learning. In *IEEE INFOCOM 2019 - IEEE Conference on Computer Communications Workshops (INFOCOM WKSHPS)*, pages 1039–1040.

Paillisse, J., Ferriol, M., Garcia, E., Latif, H., Piris, C., Lopez, A., Kuerbis, B., Rodriguez-Natal, A., Ermagan, V., Maino, F., and Cabellos, A. (2018). Ipchain: Securing ip prefix allocation and delegation with blockchain. In *2018 IEEE International Conference on Internet of Things (iThings) and IEEE Green Computing and Communications (GreenCom) and IEEE Cyber, Physical and Social Computing (CPSCom) and IEEE Smart Data (SmartData)*, pages 1236–1243.

Qingye Ding, Zhida Li, Batta, P., and Trajkovic, L.´ (2016). Detecting bgp anomalies using machine learning techniques. In *2016 IEEE International Conference on Systems, Man, and Cybernetics (SMC)*, pages 003352–003355.

Saad, M., Anwar, A., Ahmad, A., Alasmary, H., Yuksel, M., and Mohaisen, A. (2019). Routechain: Towards blockchain-based secure and efficient bgp routing. In *2019 IEEE International Conference on Blockchain and Cryptocurrency (ICBC)*, pages 210–218.

Sermpezis, P., Kotronis, V., Dainotti, A., and Dimitropoulos, X. (2018). A survey among network operators on bgp prefix hijacking. *SIGCOMM Comput. Commun. Rev.*, 48(1):64–69.

Sfirakis, I. and Kotronis, V. (2019). Validating IP prefixes and as-paths with blockchains. *CoRR*, abs/1906.03172.

Testart, C., Richter, P., King, A., Dainotti, A., and Clark, D. (2019). Profiling BGP Serial Hijackers: Capturing Persistent Misbehavior in the Global Routing Table. In *ACM Internet Measurement Conference (IMC)*.

Tony Bates, Philip Smith, G. H. (2020). Cidr report for 1 jul 20. https://www.cidr-report.org/as2.0/. CIDR Data report last accessed 1 July 2020.

Vinit, J. and Brad, E. (2018). *BGP Message*. Cisco Press, San Fransisco.

Wu, Q. and Feng, Q. (2009). *Abnormal BGP Routing Dynamics Detection by Active Learning Using Bagging on Neural Networks*, pages 61–72. Springer Berlin Heidelberg, Berlin, Heidelberg.

Xing, Q., Wang, B., and Wang, X. (2018). Bgpcoin: Blockchain-based internet number resource authority and bgp security solution. *Symmetry*, 10:408.